\documentclass[twocolumn]{aastex631}

\usepackage[english]{babel}
\usepackage[utf8x]{inputenc}
\usepackage[T1]{fontenc}
\usepackage{amssymb}
\usepackage{natbib}

 \newcommand{\scc}[1]{\textsc{#1}}

\def\bi#1{\hbox{\boldmath{$#1$}}}

\newcommand{\lya}{Ly$\alpha$}

\newcommand{\beq}{\begin{equation}}
\newcommand{\eeq}{\end{equation}}
\newcommand{\bc}{\begin{center}}
\newcommand{\ec}{\end{center}}
\newcommand{\bfig}{\begin{figure}}
\newcommand{\efig}{\end{figure}}

\renewcommand{\vec}[1]{\mathbf{#1}}

\usepackage{amsmath}
\usepackage{graphicx}
\usepackage[colorinlistoftodos]{todonotes}
\usepackage[colorlinks=true]{hyperref}

\begin{document}

\title{\scc{hyphy}: Deep Generative Conditional Posterior Mapping of Hydrodynamical Physics}
\author{Benjamin Horowitz}
\affil{Department of Astronomy, Princeton University, Princeton, NJ, USA}
\affil{Lawrence Berkeley National Lab, 1 Cyclotron Road, Berkeley, CA 94720, USA}

\email{bhorowitz@princeton.edu}

\author{Max Dornfest}
\affil{Lawrence Berkeley National Lab, 1 Cyclotron Road, Berkeley, CA 94720, USA}
\affil{Department of Physics, University of California at Berkeley, Berkeley, CA 94720, USA}
\email{Dornfest@berkeley.edu}

\author{Zarija Luki\'c}
\affil{Lawrence Berkeley National Lab, 1 Cyclotron Road, Berkeley, CA 94720, USA}

\author{Peter Harrington}
\affil{Lawrence Berkeley National Lab, 1 Cyclotron Road, Berkeley, CA 94720, USA}

\begin{abstract}
Generating large volume hydrodynamical simulations for cosmological observables is a computationally demanding task necessary for next generation observations. In this work, we construct a novel fully convolutional variational auto-encoder (VAE) to synthesize hydrodynamic fields conditioned on dark matter fields from N-body simulations. After training the model on a single hydrodynamical simulation, we are able to probabilistically map new dark matter only simulations to corresponding full hydrodynamical outputs. By sampling over the latent space of our VAE, we can generate posterior samples and study the variance of the mapping. We find that our reconstructed field provides an accurate representation of the target hydrodynamical fields as well as a reasonable variance estimates. This approach has promise for the rapid generation of mocks as well as for implementation in a full Bayesian inverse model of observed data. 
\end{abstract}

\keywords{cosmology: observations — intergalactic medium — quasars: absorption lines — galaxies: halos — techniques: spectroscopic - methods: numerical}
\section{Introduction}

Understanding the large-scale structure of the universe requires simultaneous analysis of both the evolution of the underlying dark matter cosmic web and the complex hydrodynamics leading to the formation of biased tracers. Over the past thirty years, hydrodynamical simulations have become the standard tool to generate mock observable data which includes both of these effects \citep{1990ApJ...363..349E,1992ApJS...78..341C,1996ApJS..105...19K,2005MNRAS.364.1105S}. However, the power of these hydrodynamical simulations comes with significant computational cost, and the next generation of cosmological surveys will require unprecedented precision across a wide range of scales (e.g. \citet{Walther2021}). In this regime, computing quantities like covariance matrices (which require large numbers of simulations) becomes an increasingly daunting task, so there is a clear need for approximate methods that can ease some of the computational burden.

In recent years, machine learning techniques have emerged as promising surrogate models for  complex hydrodynamics, as they can be used to rapidly generate hydrodynamic fields with remarkable perceptual and statistical fidelity.  In \cite{2019HIGAN}, the authors were able to generate realistic neutral hydrogen (HI) maps which reproduce the properties of hydrodynamical simulations over a range of scales. In \cite{2019painting}, the authors used generative models to map from two dimensional dark matter maps to thermal Sunyaev Zeldovich (tSZ) maps. They were able to reproduce accurate tSZ summary statistics over a wide range of scales, given only the dark matter maps. Related work in \cite{2020arXiv200710340W} used a more traditional feed-forward architecture, HInet, to paint neutral hydrogen in all three dimensions, but this architecture does not allow exploration of posterior properties and uncertainties. 

Estimation of the uncertainty of a neural network's output is critical in order to propagate errors accurately for cosmological and cosmographical analysis. However, within the astronomical community there has been relatively little work in error analysis in the context of these neural-network-based surrogate models. A promising approach in the case of low dimensional data is to make the network output be a multidimensional Gaussian distribution as opposed to a single point estimation, i.e. a Gaussian mixture model (see, for example \cite{2019ApJ...877L..14T}). However, for high dimensional outputs this approach would have difficulty capturing the full covariance in a memory efficient way.

In this work, we will instead structure the latent space of a conditional variational autoencoder (C-VAE) to learn the uncertainty in the mapping from a dark matter map to the hydrodynamical quantities. The general structure of our network is inspired by style transfer machine learning literature \citep{johnson2016perceptual,esser2018variational} where the latent space of the C-VAE is used to capture stylistic characteristics of the mapping. A similar network has recently been used to generate a realistic distribution of galaxy images \citep{2020arXiv200803833L}.

As we expect hydrodynamic quantities to be quasi-local, we constrain our model to maintain this property by restricting the spatial field of view of the input and use convolutional layers of different sizes in order to capture information across a range of scales. When transforming to redshift space, as is needed to match observables, this locality is broken, so rather than directly modeling observable quantities we instead focus on reconstructing the underlying (real space) baryon density, temperature, and velocity fields. Then, estimates for the target observables can trivially be computed using existing analytical tools, which also allows flexibility in modeling physical details "orthogonal" to the dark matter and hydrodynamics relationship (e.g. atomic species ratios, ionization rates, etc). For this work we will focus on \lya\ flux for Lyman Alpha Forest cosmology measurements, but our model outputs are generic and could be applied to generate other target observables. 

This work is a companion work to Harrington et al. (2021) which used a deterministic network to perform this mapping. While this approach successfully recovers the key \lya\ observables for next generation cosmological measurements, it has difficulty in capturing stochastic processes such as shocked regions. As the approach and goal of these works is quite different we present them as two separate papers. 

The paper is organized as follows: in  
\hyperref[sec:method]{Sect. 2} we briefly describe \scc{nyx} and the simulation data set used and then describe our neural network architecture. We present our results in \hyperref[sec:method]{Sect. 3}, first reviewing MAP performance and posterior accuracy. We conclude \hyperref[sec:conclusion]{Sect. 4}
with our ongoing work and future areas of interest to the community.

\begin{figure*}[ht]
    \centering
    \includegraphics[width=0.9\textwidth]{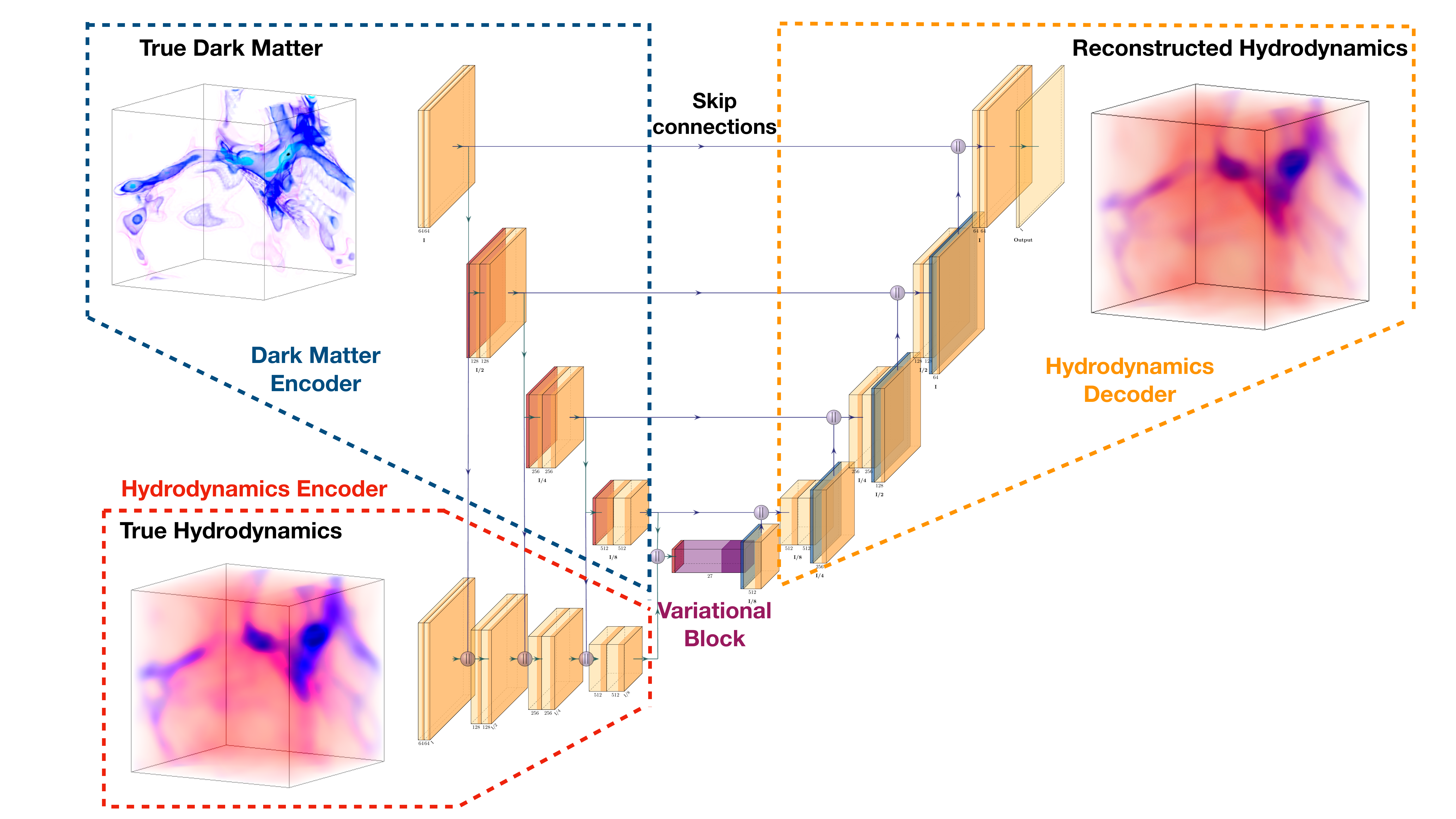}
    \caption{Schematic diagram showing the flow of our u-net conditional variational autoencoder architecture. Our model can be viewed as four interconnected parts, a encoder for the dark matter fields, an encoder for the hydrodynamical fields, a variational block, and a decoder. The network is constructed such that after training the hydrodynamical block can be removed and the latent space sampled from a unit Gaussian to generate the corresponding hydrodynamical field. Plotted is only the dark matter density and the baryon temperature, but the model also is fed as an input the dark matter velocity and outputs the baryon density and line of sight velocity.}
    \label{fig:HyPhy_train}
\end{figure*}
\begin{figure}
    \centering
    \includegraphics[width=0.48\textwidth]{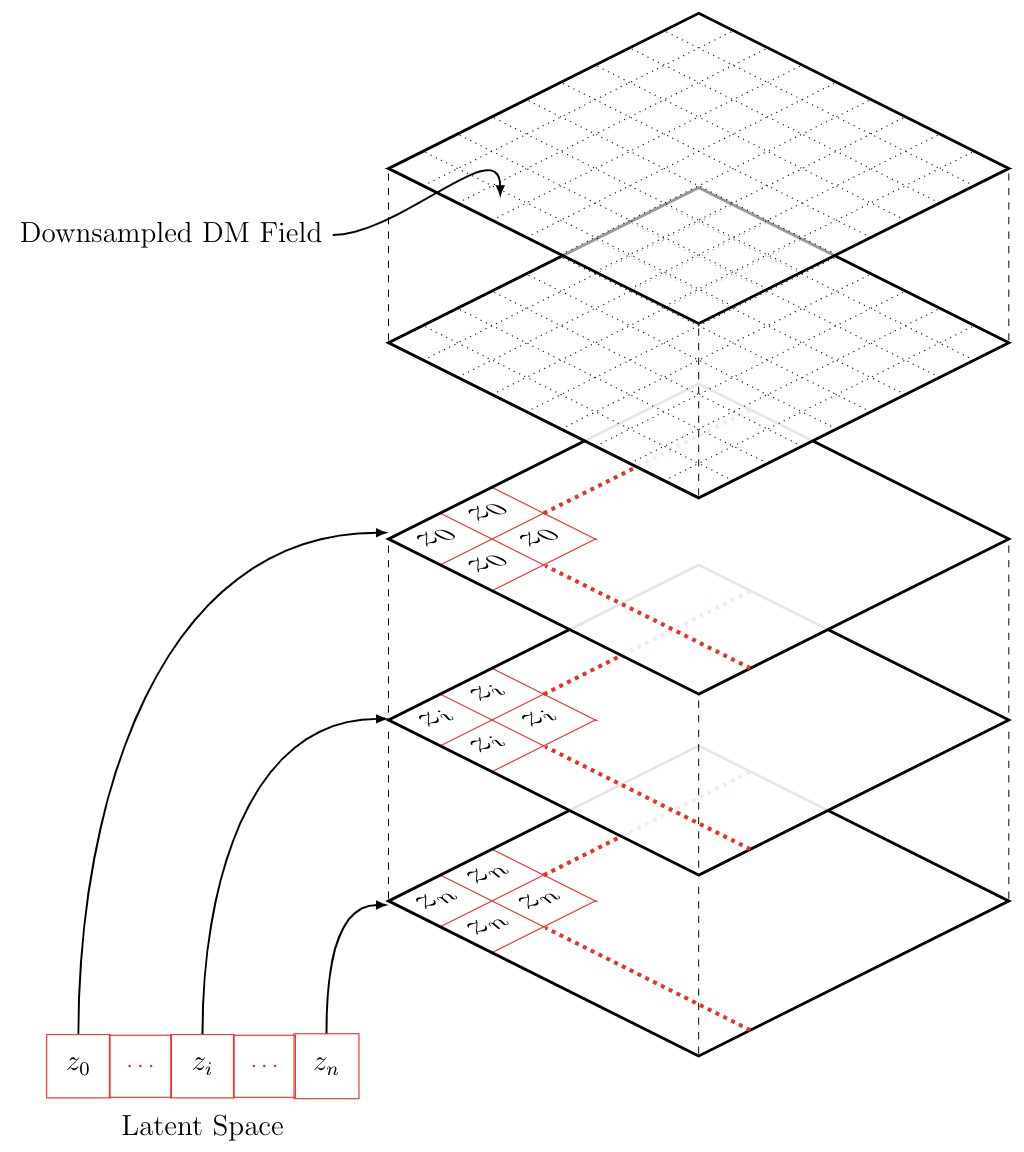}
    \caption{Diagram showing the broadcasting of the latent space samples into the same dimensionality as the downsampled dark matter field. This stack will comprise the filters passed to the upsampling convolutional layers in the hydrodynamics decoder. This allows changing input dark matter map sizes during generation while keeping the same dimensionality of the latent space and avoiding the need for dense layers. Note that in our model each filter is three dimensional, not two dimensional as shown here.}
    \label{fig:broadcast}
\end{figure}

\section{Methodology}
\label{sec:method}

\subsection{Hydrodynamical Simulations}
\label{subsec:nyx}

We choose to obtain simulation data from \scc{nyx}, a massively parallel multiphysics code, because it was developed for simulations of the intergalactic medium and has been used for many recent IGM studies \citep{2019TARDIS,davies2019,Walther2019}, and is capable of modeling dark matter and hydrodynamic evolution in great detail.
The Nyx code \citep{2013nyx} follows the evolution of dark matter modeled as self-gravitating Lagrangian  particles, while baryons are modeled as an ideal gas on a set of rectangular Cartesian grids.
The Eulerian gas dynamics equations are solved using a second-order accurate piecewise parabolic method to accurately capture shocks.
Besides solving for gravity and the Euler equations, we also include the main physical processes relevant for the Ly$\alpha$
forest.  We consider the chemistry of the gas as having a primordial composition of hydrogen and helium, include inverse Compton
cooling off the microwave background and keep track of the net loss of thermal energy resulting from atomic collisional processes \citep{2015nyxlya}. All cells are assumed to be optically thin to ionizing radiation, and radiative feedback is accounted for via a spatially uniform, time-varying 
UV background radiation given to the code as a list of photoionization and photoheating rates \citep{Haardt2012}.
We note that this type of simulation is used as a forward model in virtually any recent inference work using Lyman alpha power spectrum
\citep{Boera2019, Walther2019, PDB2020, Rogers2020, Walther2021}.
Simulations of this kind neglect the effects of inhomogeneous reionization, which produces temperature and UV background fluctuations on large scales, especially at redshifts higher than those studied in this work ($z \gtrsim 4$).

In this work we used simulations of a standard $\Lambda$CDM cosmological model, consistent with the latest cosmological constraints from the CMB \citep{Planck2020}: $\Omega_{\rm m}$=0.31, $\Omega_\Lambda$=0.69, $\Omega_b$=0.0487, $h$=0.675, $\sigma_8$=0.82 and $n_s$=0.965. For the hydrogen and helium mass abundances we adopted values consistent with the CMB observations and Big Bang nucleosynthesis \citep{Coc2013}: $X_p$=0.76 and $Y_p$=0.24.
Box size of simulations is $20$ $h^{-1}$Mpc, with $N = 1024^{3}$ particles and grid cells, resulting in a resolution of $\sim 20 h^{-1}$ kpc, fulfilling convergence criteria for a percent-level accurate Ly$\alpha$ quantities \citep{2015nyxlya}.

In addition to hydrodynamical simulations, we have also produced N-body simulations starting with the same initial conditions. These neglect all other forces but gravity, and all matter is considered collisionless, although baryonic effects are imprinted in the initial power spectrum for the total matter.  Baryons only minorly affect dark matter clustering in the regime relevant for the Ly$\alpha$ forest, but we have nevertheless produced these ``companion'' simulations to maintain maximum reality in our modeling when we train to infer hydrodynamical quantities from a N-body run and avoid any possible back-reaction on the dark matter field.  Throughout this work we use one set of hydro and matching N-body simulation for training and a different set for validation purpose.  The two differ only in the random phases of the Gaussian random field in the initial conditions.





\begin{figure*}
    \centering
    \includegraphics[trim=1cm 2.5cm 1cm 2.5cm,width=0.95\textwidth]{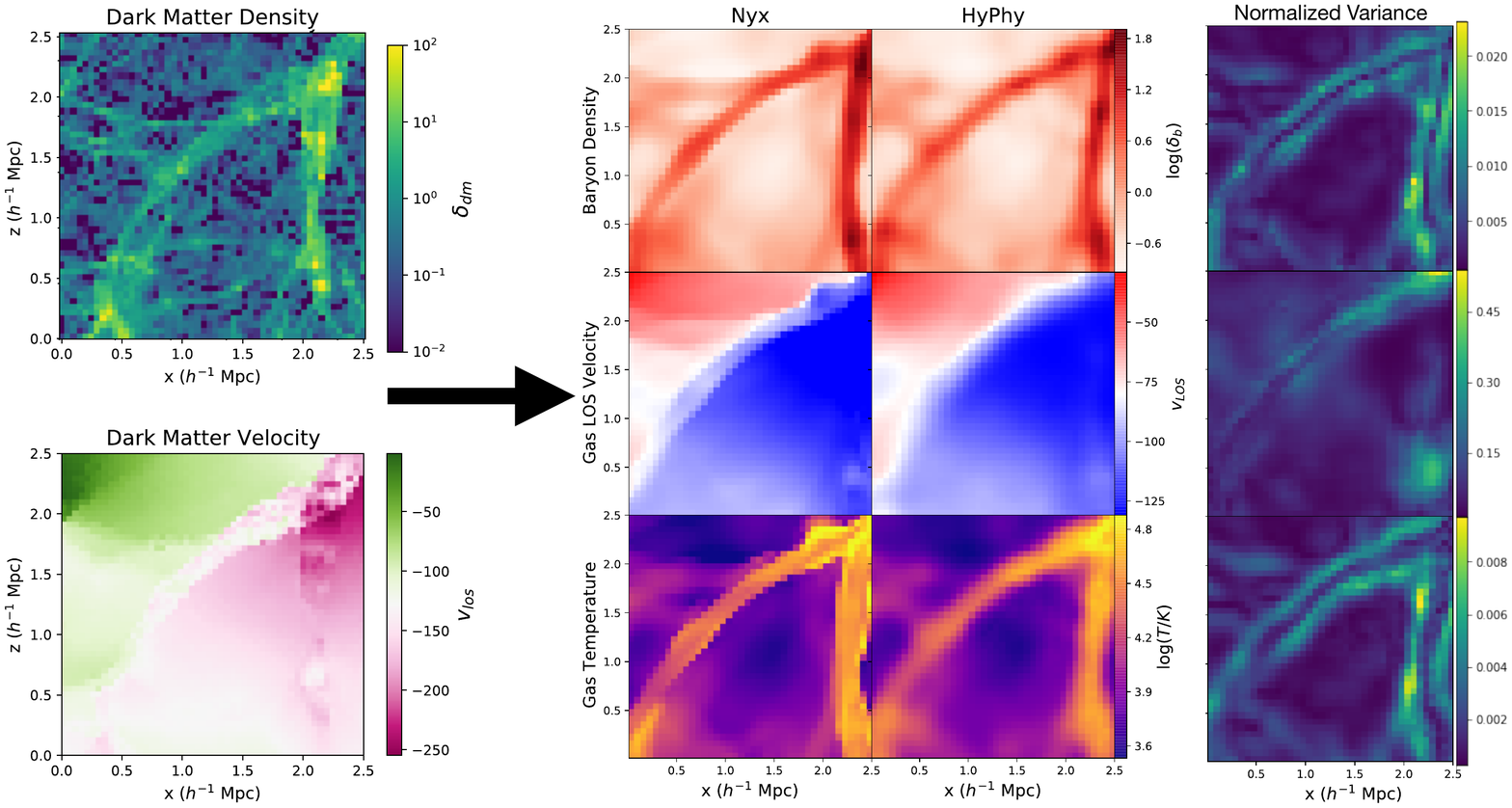}
    \caption{A typical example single slice of the \scc{hyphy} mapping is shown. On the left are the test input dark matter only maps and in the center are the hydrodynamical fields (baryon density, velocity, temperature) are compared. There is strong qualitative agreement, with the network accurately learning various characteristics of the hydrodynamical fields including the variable baryon pressure smoothing and thermal properties. On the far right the networks' estimated variance is shown, calculated from 1000 samples over the latent space, normalized by the mean value of the field.}
    \label{fig:sixpanel_noshock}
 \end{figure*}
 
\begin{figure*}

    \centering
    \includegraphics[trim=1cm 2.5cm 1cm 2.5cm,width=0.95\textwidth]{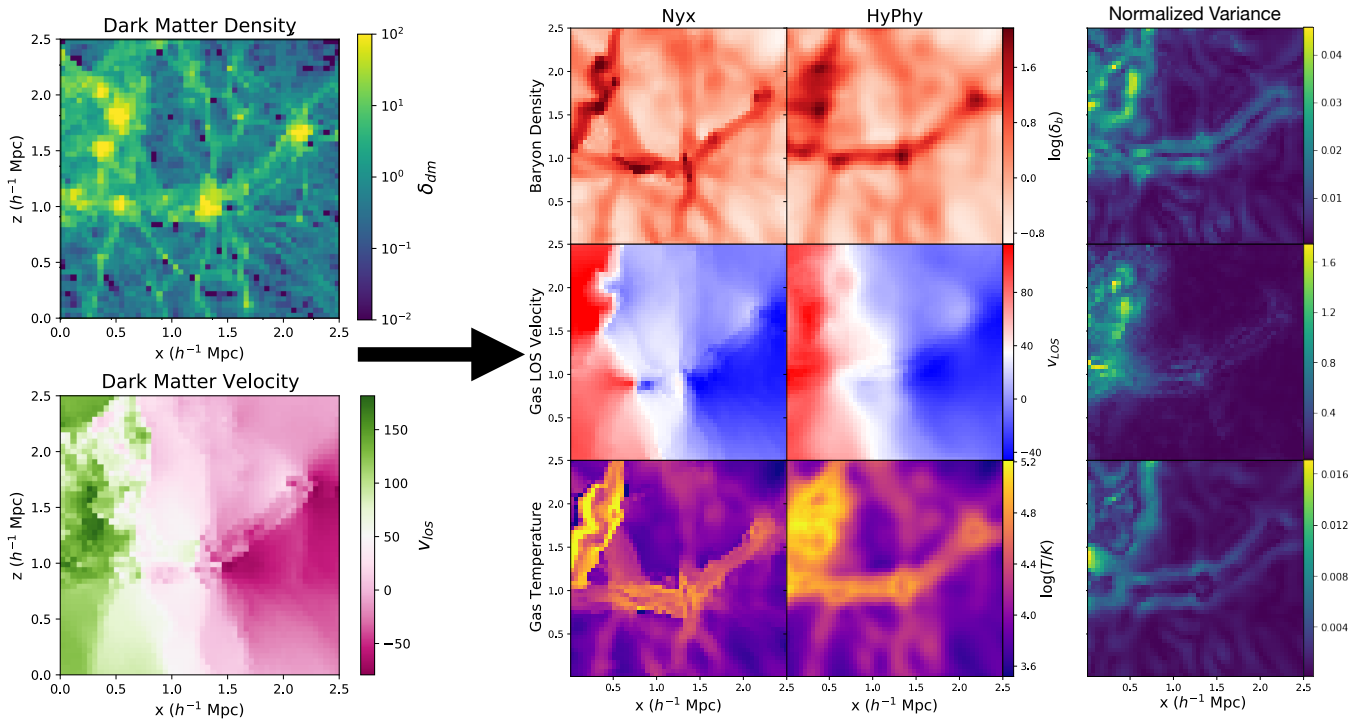}
    \caption{Same as Fig \ref{fig:sixpanel_noshock}, but showing one of the worse reconstructed test boxes. The characteristics of moderately large shocks are difficult for our network to learn, resulting in a less sharp feature. However, this uncertainty in the reconstruction is captured when sampling over latent space, as we see significant variance in the shocked region. If we look at the individual posterior samples this is even more visually apparent, see Figure \ref{fig:shock_samples}.}
    \label{fig:sixpanel_shock}
\end{figure*}
\subsection{Conditional Variational Auto-encoder Model}
\label{subsec:nn}

As neural networks are becoming an increasingly well-known tool in astrophysics and cosmology, we just briefly summarize our model\footnote{Our model is implemented in \scc{tensorflow}, with sample data and a saved model available at \url{https://github.com/bhorowitz/HyPhy-public}.} and highlight how it differs from others in the literature. 

We use a conditional variational auto-encoder architecture to study the posterior of hydrodynamical quantities, $\tau$, given the dark matter field, $\delta$.  The core concept of a variational auto-encoder network is that the neural network learns the probability space described by the training sample by marginalizing over a (usually bottlenecked) set of latent space parameters, $Z$. In this case we are interested in constraining the outputs of our model to those corresponding to a given dark matter realization which we enforce by using the dark matter realization as a ``condition"; a field given to the network both during the ``encoding" step (where the fields are mapped to the latent space) and the ``decoding" step (where the latent space is mapped to a 3D field). For finding a given hydrodynamical realization, we are interested in calculating the following quantity,
\begin{equation}
    P(\tau | \delta) = \int p(\tau|Z, \delta) p(Z | \delta) dZ,
\end{equation}
where $p(\tau|Z,\delta)$ is the generator network. In order to train the network, we also need to define a connected, overlapping, encoder network, $q(Z| \tau, \delta)$. We can generalize the standard evidence lower bound (ELBO) straightforwardly to include this conditioning variable
\begin{eqnarray}
      \log P(\tau | \delta)=& \log \int p(\tau|Z, \delta) p(Z | \delta) dZ \geq \mathbb{E}_q \left[\log \frac{p(\tau,\delta | z)}{q(Z|\tau,\delta}) \right]\nonumber \\
      =& \mathbb{E}_q  \left[\log \frac{p(\tau |\delta, Z) p(Z|\delta)}{q(Z|\tau,\delta)}\right],
\end{eqnarray}
where $\mathbb{E}_q$ is the expectation value marginalized over $q$. In standard style-transfer implementations, $p(z|\delta)$ plays an important role as a prior over the latent space parameters (i.e. \cite{esser2018variational}). In our case, we will not be imposing any direct interpretation to our latent space parameters, and we will find that this prior distribution, $p(Z|\delta)$, will be pulled to a unit-normal Gaussian due to the loss term.  

We again follow the standard derivation for CVAE networks and model the probability distributions, $p(\tau | \delta, Z)$ and $q(Z|\delta, \tau$), as neural networks with associated free parameters, $G_\theta$ and $F_\phi$, respectively. In Figure \ref{fig:HyPhy_train}, $G_\theta$ corresponds to the combination of the dark matter encoder and hydrodynamics decoder, while $F_\phi$ corresponds to the combination of dark matter encoder and hydrodynamics encoder. These overlapping neural networks weights, $\theta$ and $\phi$, are trained jointly using the standard ADAM optimizer under the associated loss
\begin{eqnarray}
    \mathcal{L}(\delta,\theta,\phi) &= - \textrm{KL} (q_\phi(Z|\delta,\tau)||p_\theta(Z| \tau)) \nonumber\\ &+ \mathbb{E}_{q_\phi(Z|\delta,\tau)}\left[ \log p_\theta(\tau| Z, \delta) \right],
\end{eqnarray}
where $\textrm{KL}$ is the Kullback-Leibler divergence to compare distributions. Assuming we treat the generator network, $G_\theta$, as deterministic in $\delta$ and $Z$, we can simplify the second term with our chosen reconstruction loss. There are many possible choices for this loss function, including $L1$, $L2$, perceptual-loss, or an adversarial loss. In our case we choose the standard $L1$ loss term, i.e. 
\begin{equation}
    \mathcal{L}(\delta,\theta,\phi) = - \textrm{KL} (q_\phi(Z|\delta,\tau)||p_\theta(Z| \tau)) + || \hat{\tau} - G_\theta(\delta,Z)||_1,
\end{equation}
where $\hat{\tau}$ is the simulated true field corresponding to $\delta$. The KL-divergence term can also be further simplified by expressing our latent space image, $q_\phi(Z|\delta,\tau)$, as a multidimensional Gaussian with diagonal covariance using the renormalization trick and our target distribution as a unit normal Gaussian,
\begin{eqnarray}
    \mathcal{L}(\delta,\theta,\phi) &= - \textrm{KL} (\mathcal{N}(\mu(\tau,\delta),\sigma(\tau,\delta))||\mathcal{N}(0,1))\nonumber\\  &+ || \tau - G_\theta(\delta,z)||_1,
\end{eqnarray}
where $\mu$ and $\sigma$ are outputs of our encoding network $F_\phi$, and correspond to the means and standard deviations of our latent space distributions from which we sample.


The variations of the hydrodynamical model are captured by properties of the dark matter field at a variety of scales. We are therefore motivated by examples of image segmentation from machine learning literature to use an altered U-Net structure for our network. We summarize this structure in Figure \ref{fig:HyPhy_train}. The most notable in the structure of this network is the skip connections across the bottleneck mapping the dark matter field from the encoder to decoder side. This is designed to maximize the possible information extracted from the dark matter field, in a computationally expedient fashion, as well as to minimize the dark matter dependent structure in the latent space. In particular since we draw samples from the latent space that are fully representative of uncertainty in the mapping, we want the conditional probability of $p(Z | \delta)$ to be as close to a uniform Gaussian, $\mathcal{N}(0,1)$ as possible. Structure in the latent space will result in a biased posterior sample distribution. In abstract we could implement various techniques (such as an autoregressive flow, as in \cite{2020arXiv200803833L}), to ensure that our latent space is well mixed, but in practice we find that there is little dependency on matter properties in the latent space.

\subsection{Fully Convolutional Architecture}

A critical feature of our architecture is for it to be fully convolutional, thereby allowing inputs of any size during inference while being trained on smaller volumes. Traditionally, VAEs utilize dense layers to upsample the drawn latent space variable, which are then possibly followed by transposed convolutions (or other upsampling convolutions) to reach the desired output size. This design restricts the input image size to always be of the same dimension as the training set. To avoid this constraint, we broadcast the latent space parameters into a feature map whose dimension matches that of the lowest level in the U-Net, then upsample from there as is standard (see Figure \ref{fig:broadcast}). Since the upsampling convolutional layers are inherently local (as determined by their kernel size) we maintain the desired locality of our network, while still allowing every element of the output to ``see'' the full latent space.

\begin{figure*}[t]
    \centering
    \includegraphics[trim={3.5cm 0 4cm 0},clip,width=0.95\textwidth]{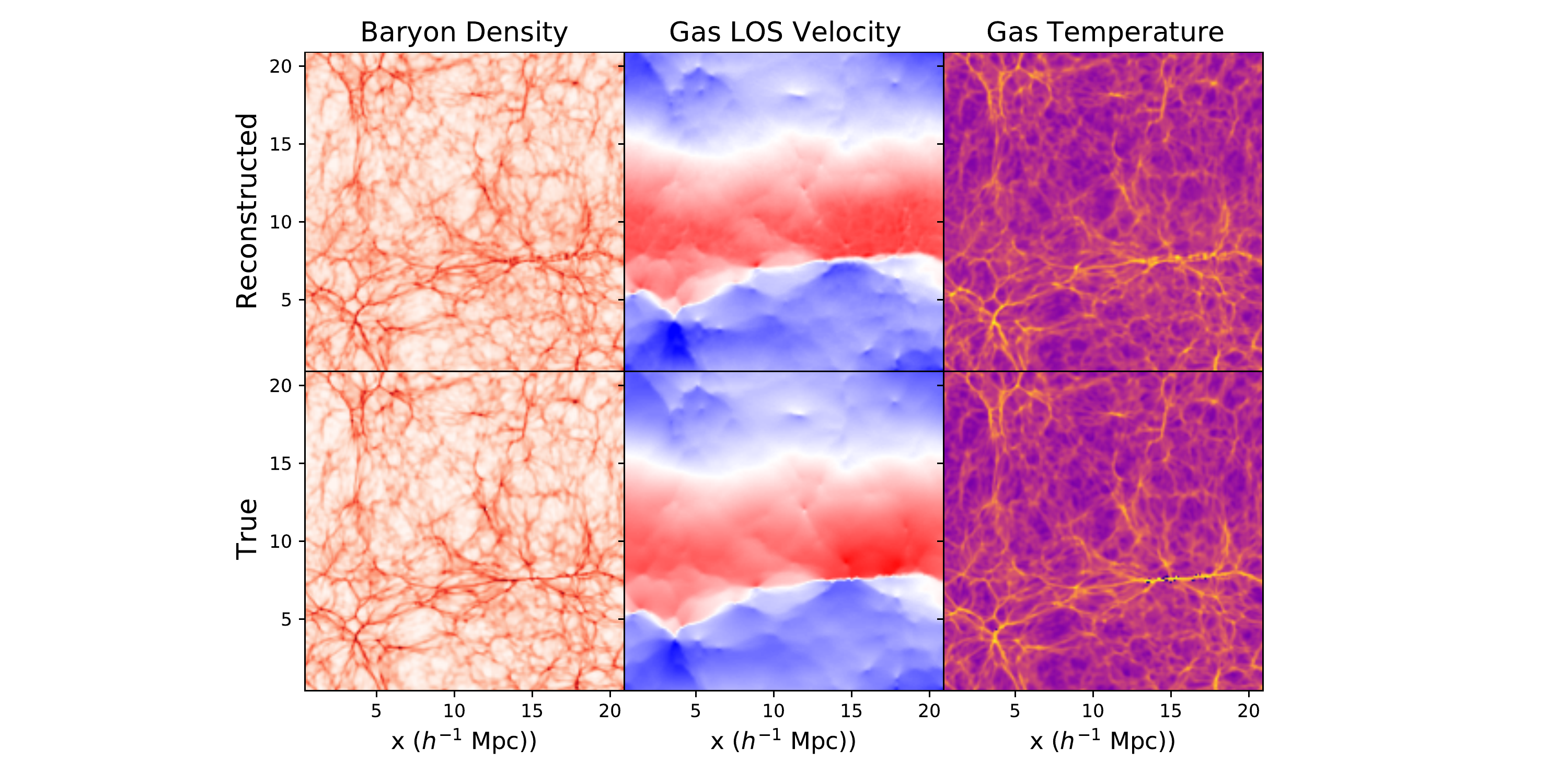}
    \caption{\scc{hyphy} run on a box significantly larger than the training volumes. The fully convolutional nature of the network allows the network to be run on any box that is a multiple of a base dimension (currently 8 pixels) and has the same spatial resolution. There are no obvious artifacts caused by the increased image size.}
    \label{fig:big_box}
\end{figure*}

While generated volumes can be of any dimension, the training volumes are fixed in size due to the need for dense layers in the encoder to predict the latent space distributional parameters. However, training data can easily be segmented to the proper size, and the choice of the crop size is a problem-dependent hyperparameter to be tuned. Training on boxes of limited volume means long distance correlations not captured in the dark matter distribution (e.g. a spatially fluctuating UV background) would not be well reproduced with this architecture and would require additional considerations. For the purpose of this work, we use training boxes of approximately 4 $h^{-1} Mpc$.

Since neither each individual training box or standard convolutional layer implementations are periodic, we must minimize edge effects by restricting our loss function to only compare the central region of our training sample. In this work, we train on boxes with $64$ voxel side-length, of which we ignore 10 voxels on each side in order to avoid dealing with edge effects.

In order to better utilize our limited training data, we use the symmetries of our system to randomly augment training samples by performing reflections and rotations of our box over each of the three spatial dimensions. In addition, our training boxes are overlapping in space, providing some knowledge of translational symmetry of the underlying physics model.

\subsection{Redshift Information}

For application to real \lya\ forest data, a key aspect of the mapping is to include the redshift dependence of the mapping as this dependence is used for cosmological constraints (i.e. \cite{2019JCAP...07..017C}). To include this in our model, we condition on a redshift field of the same size as our training volume (i.e. every pixel has an associated redshift). This allows us to vary the redshift over the box (i.e. to generate light-cones), as well as train the same model to work across cosmic time. We train our model using snapshots from the same simulation at $z=2.4$, $z=3.0$ and $z=4.0$, as this range dominates the cosmological \lya\ forest signal \citep{Walther2021}. 

\section{Results}
\label{sec:results}

We apply our trained network to a separate simulation (not used for training), focusing on the central redshift of $z=3.0$, which requires only changing the latent space projection dimension while maintaining the same network weights. We first show the results for the maximum a posterior point predicted by the network, both in the base hydrodynamical quantities and in terms of derived \lya\ flux. We then discuss the posterior properties of a representative sample distribution in Subsection \ref{subsec:posterior}.

    
    
    

\begin{figure*}
    \centering
    \includegraphics[width=0.99\textwidth]{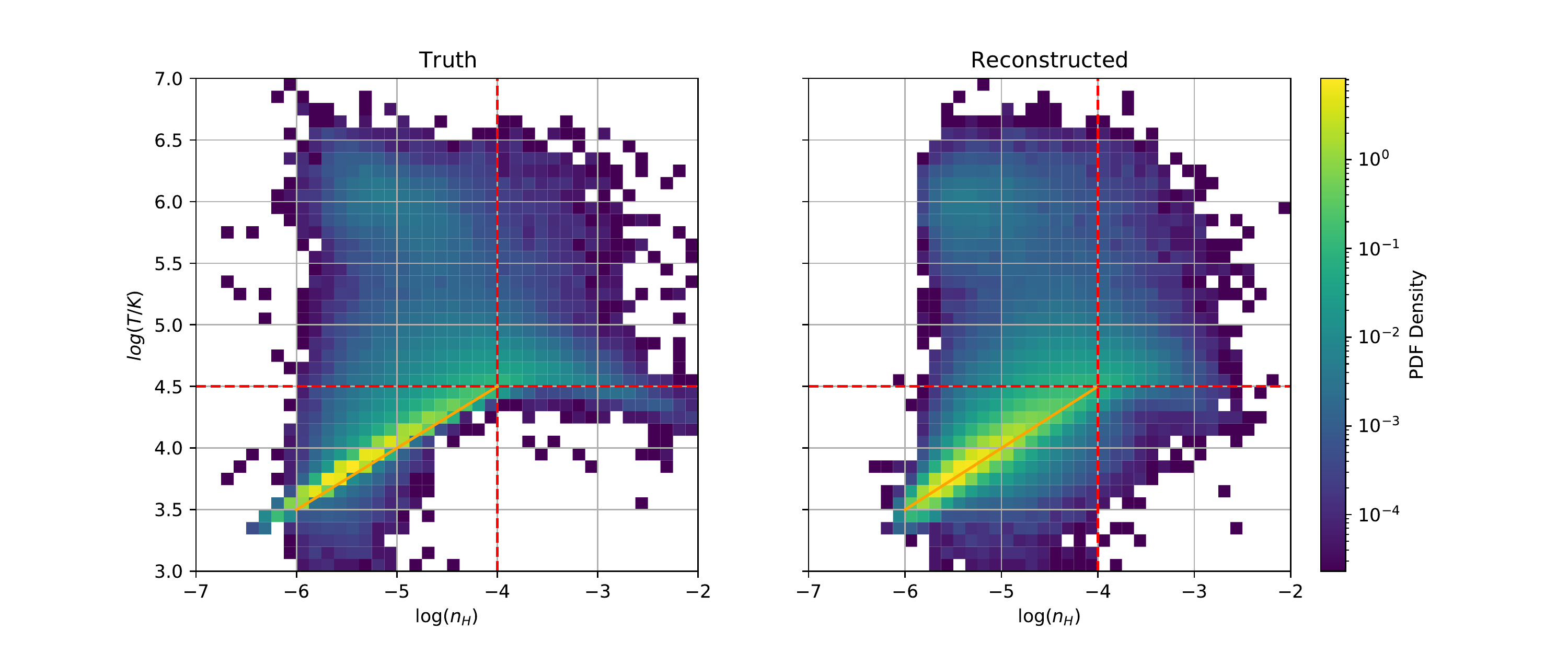}
    \caption{2D histogram of temperature and hydrogen density relation in truth compared to the reconstruction. The orange line indicates a best fit power-law solution, $T = T_0 n_H^\gamma$, across the range of $10^{-4} < n_H < 10^{-6}$. Red dashed lines indicate the classification of the gas content of the universe into wCGM, WHIM, diffuse gas, and halo gas. We find excellent agreement in terms of the mean relation, as well as the warm hot intergalactic medium component. However the cool highest density regions, corresponding to centers of cluster regions (i.e. CGM regions), are not well reconstructed.}
    \label{fig:hist2d}
\end{figure*}

The maximum a posterior (MAP) should correspond to the highest maximum of the distribution $P(\vec{Z})$, which by construction will be at $\vec{Z}=\vec{0}$ for a fully trained model. A useful starting point for our analysis will be to see how this point in the posterior space behaves to judge the quality of our reconstruction for test boxes with the same dimension as our training set. Note that we expect these reconstructions to be slightly smoothed versus generic samples.

We show test boxes both without and with significant shocked regions in Figures \ref{fig:sixpanel_noshock} and \ref{fig:sixpanel_shock}, respectively. In low to medium density regions, we recover excellent qualitative agreement across a range of scales for baryon density, baryon velocity, and temperature. For the shocked example, Figure \ref{fig:shock_samples}, the \scc{hyphy} recovered temperature field has a significantly less prominent shocked region with a much smoother boundary; this is a common phenomenon VAE-type networks \citep{2018arXiv180410323K}. 

We also show the model predicted variance by examining 1000 samples drawn from $\mathcal{N}(0,1)$, qualitatively showing that regions of highest variance are where there is the strongest disagreement between \scc{hyphy} output and the simulated true field. In particular, regions near the boundaries of structures (like filaments) have significant variance as do those with significant astrophysical shocks. We discuss these properties more in Section \ref{subsec:posterior}.

\subsection{Large Volume Statistics}

Next, we apply \scc{hyphy} to the entire test N-body simulation at once. 
To match the resolution of our training sample, we uniformly down-sample the volume by a factor of two. We show the reconstructed resulting baryon density, velocity, and temperature in Figure \ref{fig:big_box}. 



\subsubsection{Gas Phase Physics}

A well studied aspect of cosmological hydrodynamical simulations is the relation between the gas density and gas temperature \citep{Gunn1965, 2016Sorini}. We show our reconstructed relation in log space in Figure \ref{fig:hist2d}. Following \cite{2010ApJ...721...46U,2019MNRAS.486.3766M, 2020arXiv201015139G}, this plot can be viewed as a phase-space distribution between Warm Hot Intergalactic Medium (WHIM), warm Circumgalactic Medium (wCGM), diffuse IGM, halo gas, and ``hot gas". Halo gas consists of relatively ``cool" gas including ISM within galaxies, as well as more diffuse gas found in-between galaxies within halos. Similarly, warm CGM is found in dense enviroments, but has been significantly heated via shocks or feedback processes near galaxies. Diffuse IGM and WHIM are found in less dense environments, such as regions surrounding filaments. The WHIM component is of significant interest due to the ``missing baryon" problem \citep{1998ApJ...503..518F}. The  last component, consisting of gas at any density at a temperature in excess of $10^7$ K and generally associated with massive shocks, is a vanishingly small percentage of the test volume (68 pixels out of $512^3$ total pixels) and is grouped with wCGM or WHIM depending on density. In addition, some studies (e.g. \cite{2019MNRAS.486.3766M}) separate star-forming gas, with densities $\log(n_H) > -1.0$, as a separate phase. Since \scc{nyx} does not model star formation we do not include this phase in our analysis. 

In Figure \ref{fig:hist2d}, we summarize the recovered volume fractions vs. the simulated truth. We find excellent recovery of the diffuse IGM and WHIM, with slightly worse performance of the wCGM component. Halo Gas, constituting a tiny fraction of the total volume, is very difficult to recover accurately, with \scc{hyphy} finding a factor $19\times$ more halo gas. However, it is important to note the vast majority of this excess is on the border of the diffuse IGM component in phase space and can probably be better described as over-estimated IGM temperature as opposed to misattributed halo gas.  

A key property of interest to the IGM is the power law relation between density and temperature, which can be related to statistics of the \lya\ forest \citep{1997MNRAS.292...27H}. Following the procedure in \citet{2016Sorini}, we identify the gas around two bins centered at $\log(\Delta_{b,0}) = -1$ and $\log(\Delta_{b,1}) = 0$, with a width of $5\%$ around the central value. We calculate the median temperature of the corresponding particles in each bin, and use those two points to determine the power-law relation $T = T_0 \Delta_b^\gamma$. For \scc{nyx}, we find $(T_0, \gamma) = (10^{4.08} K, 1.53)$ and for the \scc{hyphy} reconstruction we find $(T_0, \gamma) = (10^{4.08} K, 1.51)$. 





\subsection{Posterior Exploration}
\label{subsec:posterior}

The main motivation of our architecture is to allow accurate unbiased posterior sampling of our hydrodynamical quantity through Gaussian sampling of our latent space variable. In this subsection we explore how accurate this mapping is.

\begin{figure}
    \centering
    \includegraphics[width=0.450\textwidth]{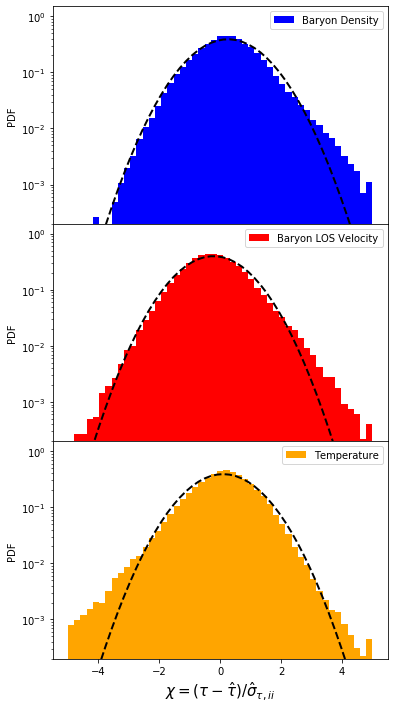}
    \caption{Distributions of the $\chi$ value for each of our three reconstructions. Also plotted is a unit normal Gaussian distribution, showing close agreement. Excess values along the tails indicate model failure, either to estimate proper variance or strong residual biases, or potentially importance of off-diagonal covariance terms not captured in this analysis.  We find these value points constitute a very small volume fraction ($\lesssim 0.1\%$). The field with the most variance that is not captured by our model is the temperature field, which is also the most numerically complex to calculate in the original hydrodynamical simulation due to shock physics.}
    \label{fig:hist_chi}
\end{figure}

This question is not straightforward as it is computationally difficult to calculate the true posterior for the target mapping, which would essentially require running millions of additional hydrodynamical simulations. With this in mind we are left to check if the samples drawn have the correct statistical properties of a posteriori sample, as well as examine its qualitative behaviors. We draw 1000 random samples from a unit Gaussian distribution in latent space and predict the hydrodynamical quantities associated with a test dark matter field. 
To test whether or not our variance estimates are accurate, we plot the $\chi$ values for our test images in Figure \ref{fig:hist_chi}. For a true posterior we would expect this distribution to be approximately Gaussian.

In particular, we can use the samples to construct a covariance matrix to use to test the statistical significance of deviations away form the MAP solution using the standard chi-squared formula,
\begin{equation}
    \chi^2 = (\bi{\tau}-\bi{\hat{\tau}})^T \bi{C}^{-1} (\bi{\tau}-\bi{\hat{\tau}}) \rightarrow (\bi{\tau} - \bi{\hat{\tau}})^2/\sigma^2_{\tau}
\end{equation}
where in the last arrow we assume a diagonal covariance for implementation/memory reasons. If each sample is truly independent and represents the posterior, we expect that the corresponding chi-squared values from the ensemble should be Gaussian distributed with zero bias and standard deviation of one. This is arguably a necessary, but not sufficient, condition for the samples to represent a maximum a posteriori sample. We show this distribution in Figure \ref{fig:hist_chi}.

\begin{figure*}
    \centering
    \includegraphics[width=0.95\textwidth]{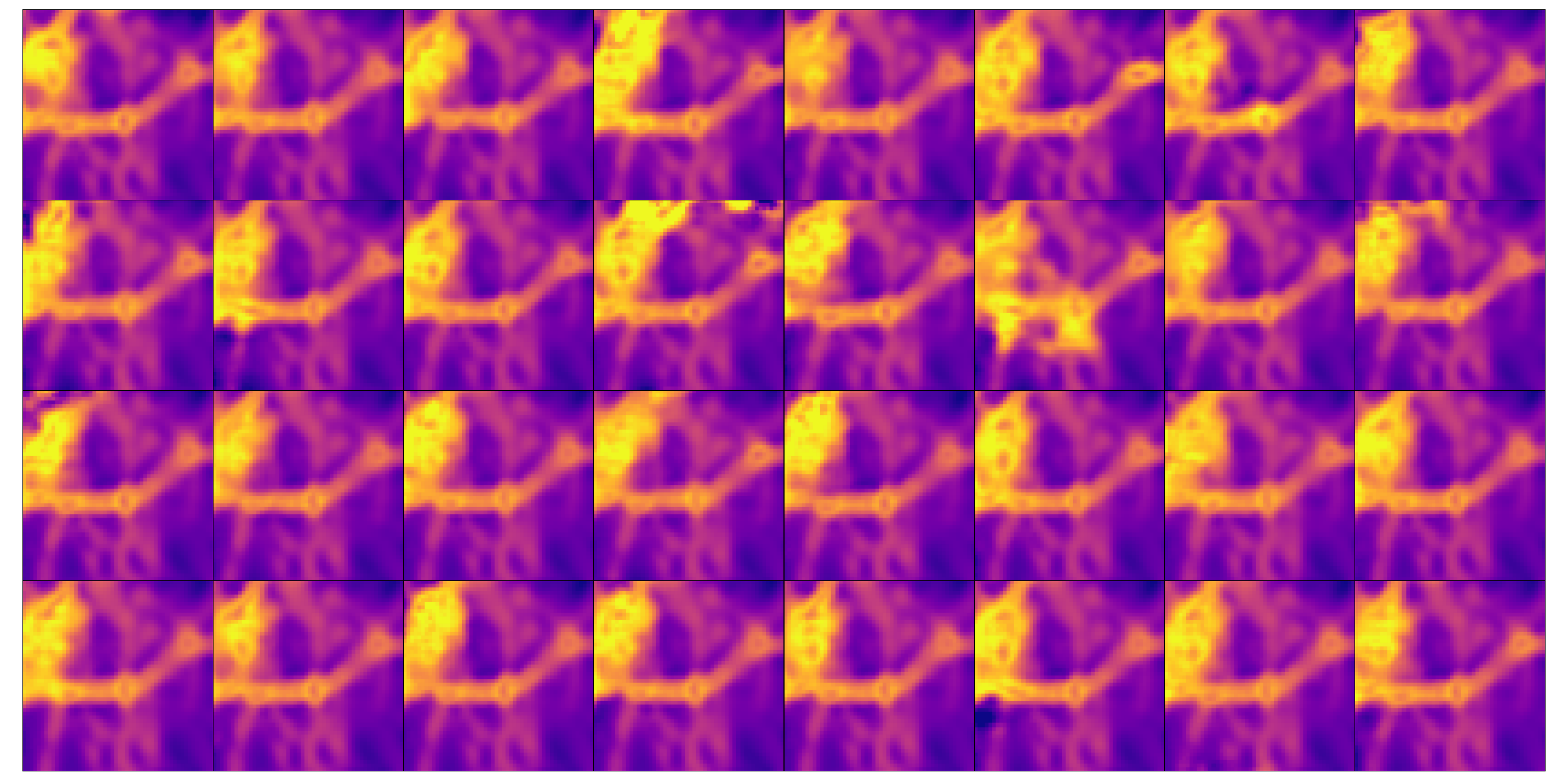}
    \caption{Postage stamp posterior samples of the temperature field from Figure \ref{fig:sixpanel_shock}. There is high variation in the shocked region, showing the hydrodynamical uncertainty the model has learned, while the areas outside the shocked region are stationary (with some occasional edge effects).}
    \label{fig:shock_samples}
\end{figure*}
An additional property desired of the posterior is for the samples to capture the qualitative uncertainty of shocked regions when sampled over latent space. We show one example of such a region in Figure \ref{fig:shock_samples}. We expect dense regions at nodes of the cosmic web to have the largest hydrodynamical effects, while those in under-dense environments away from cosmic structures should follow roughly power-law distributions without uncertainty. In the samples in Figure \ref{fig:shock_samples}, one sees high variability of the specifics of the shocked region, indicating the model is learning to account for hydrodynamical uncertainty. Again, it is difficult to formulate a rigorous method to test whether this variability is the ``true" uncertainty without running a large suite of hydrodynamical simulations.

\subsection{Predicted Lyman Alpha Forest}
\label{subsec:lyastat}

\begin{figure}%
\centering
		\includegraphics[width=3in]{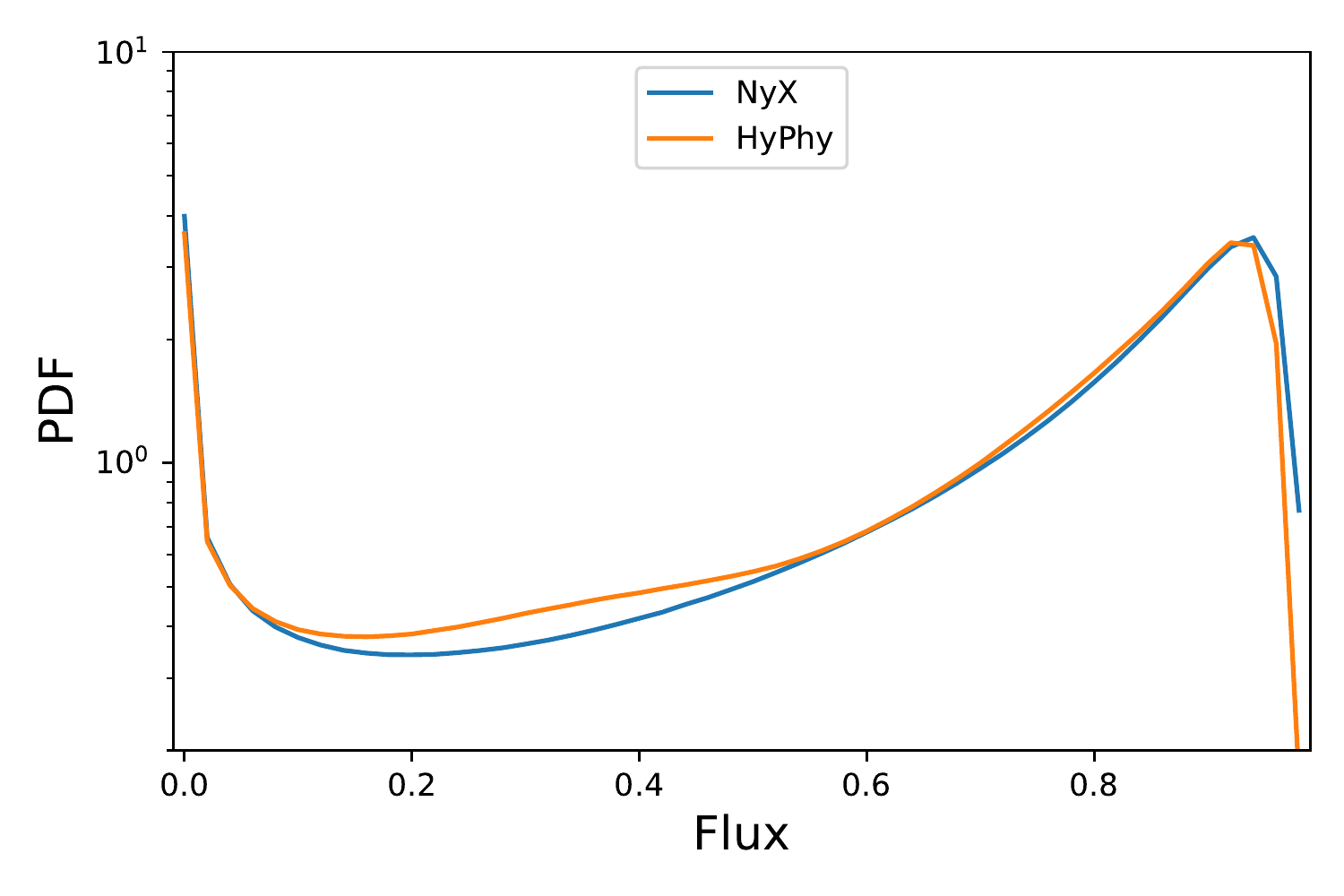}
		\caption{Performance of \scc{hyphy} in terms of the Lyman Alpha forest flux in redshift space using Equation \ref{eq:tau} as implemented in \scc{GIMLET}. We show the resulting PDF of \lya\ flux, showing substantial biases in the reconstructed flux at $F\sim 0.25$, but good agreement at the extremes.}\label{subfig:flux}		
		\centering
		\includegraphics[width=3in]{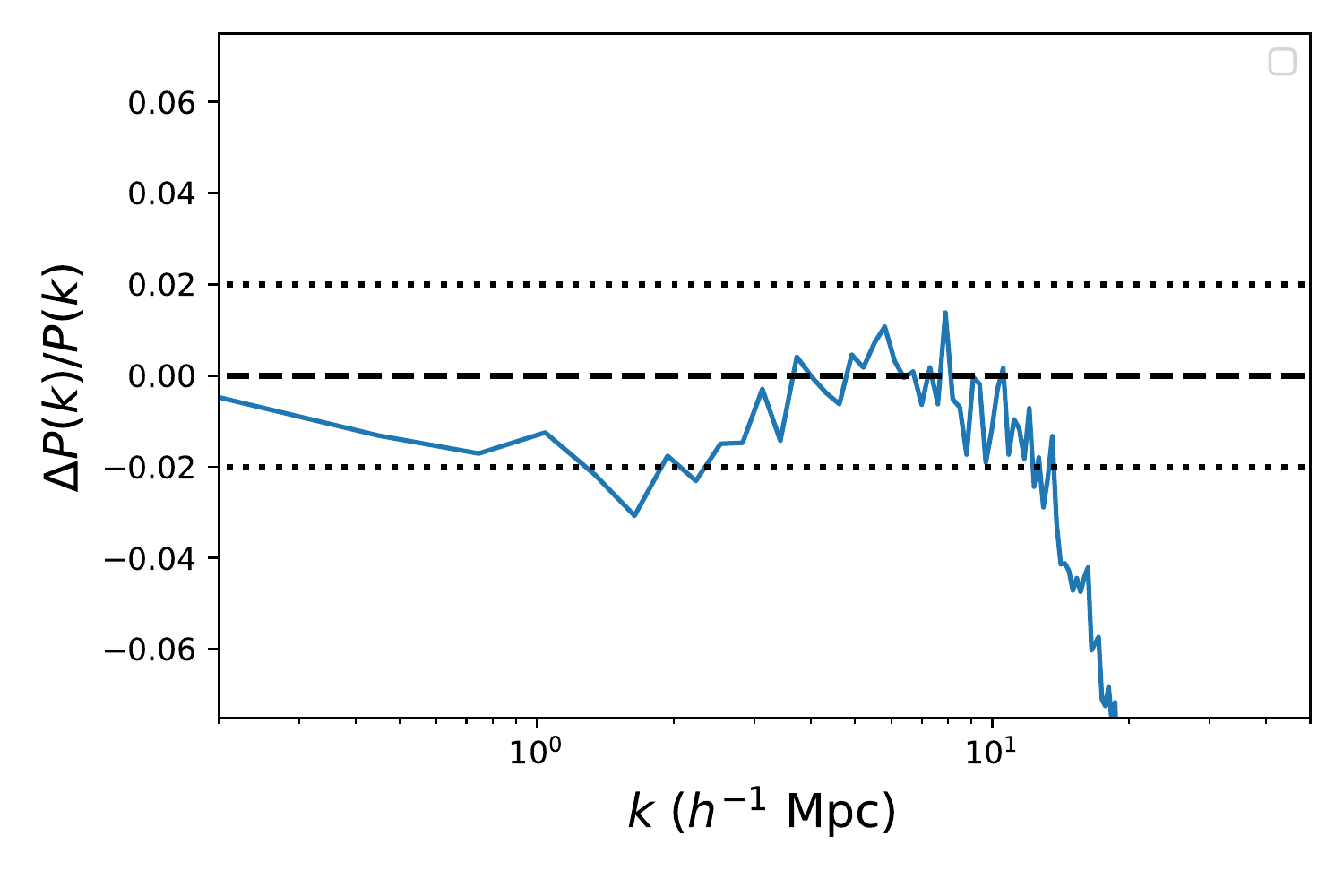}
		\caption{Performance of \scc{hyphy} in terms of the Lyman Alpha forest power spectrum in redshift space using Equation \ref{eq:tau} as implemented in \scc{GIMLET}. We show the resulting power spectra of \lya\ flux, finding good agreement across a range of scales up to $k\sim 14$ $h^{-1}$ Mpc.}\label{subfig:ps}
\end{figure}

The \lya\ forest arises from the scattering of photons at the characteristic rest frame \lya\ frequency along their path from a background source, generally either a quasar or galaxy, to the observer. The fraction of the transmitted flux is given by $F=\exp(-\tau)$, where $\tau$ is the optical depth of the intervening gas. 
The optical depth in redshift space at a given velocity coordinate $u$ along the line of sight is given by
\begin{equation} \label{eq:tau}
	\tau (u) = \int  du' \, \frac{\lambda_{\mathrm{Ly}\alpha} \sigma \, n_{\mathrm{H I}}(\boldsymbol{u'})}{H(z) b(\boldsymbol{u'})} \exp\left[-\frac{(u-(u' + u_{\mathrm{pec}}(u')))^2}{b(\boldsymbol{u'})^2}\right],
\end{equation}
where $u'$ is the component of the Hubble flow velocity field $\boldsymbol{u'}$  along the line-of-sight, over which the integral is calculated and  $b(\boldsymbol{u'})=\sqrt{2 k_B T(\boldsymbol{u'})/m_p}$ is the thermal broadening of the absorption feature. With the output of the \scc{hyphy} model, we have all the components necessary to calculate the predicted \lya\ forest statistics. We run the \scc{gimlet} \citep{2016ComAC...3....4F} library to numerically calculate Eq. \ref{eq:tau} on both the original simulation and our predicted output, using a mapping from hydrogen fraction to neutral hydrogen from \citet{2013MNRAS.430.2427R}. Flux distributions are calculated along each of three axes and then averaged out.  In Figure \ref{subfig:flux}, we show the resulting flux distribution and in Figure \ref{subfig:ps} we show the error on the reconstructed power-spectra. 


\section{Conclusion}
\label{sec:conclusion}

In this work, we have provided a flexible generic mapping from dark matter fields to hydrodynamical quantities, in particular the gas temperature, density, and velocity. We demonstrated that this mapping provides accurate reconstructions across a wide range of scales and captures a number of the statistical properties of the underlying truth fields. In addition to constructing this mapping, the underlying model provides posterior samples which are consistent with the underlying dark matter field. While it is computationally infeasible to analytically calculate the true maximum likelihood posterior in this case, we show that our posterior samples are a valid posterior through their variance properties.

It is important to note that we were able to construct these mappings despite only training on a single \scc{nyx} simulated run, using data at $z=2.4$, $z=3.0$, and $z=4.0$. This was possible due to various data augmentations used which exploit the symmetry of the hydrodynamical physics as well as a regularizing effect of the underlying CVAE architecture \citep{kamyab2019deep}. Despite this relatively small training volume, we are still able to capture the qualitative effects and associated uncertainties in shocked regions. 

We do find some cosmic environments where there are constant residual biases which, despite significant testing, we were unable to reduce completely. While we recover the statistical properties of the WHIM, WCGM and diffuse IGM very well, recovering the hot halo gas contribution is significantly more difficult. We find that our model systematically under-estimates the density in these regions, while overestimating its overall volume fraction, resulting in an incorrect estimate for the phase space distribution. Gas in this region accounts for a very small small percent of the total volume, $\sim 4 \times 10^{-6} \%$, likely resulting in \scc{hyphy} having difficulty capturing its properties. 

For \lya\ Forest analysis, these hot CGM regions do not have a noticeable effect on the forests' statistical properties so we do not focus on optimizing this aspect. Going beyond this work, one could perform various data augmentations to increase its importance in the loss to result in better model performance in this region. Approaches to dealing with such unbalanced data are well studied in the machine learning literature \citep{wang2016training}, including oversampling the minority data \citep{khoshgoftaar2007learning} and using Generative Adversarial Networks to create synthetic minority data \citep{2019arXiv190409135K}.


One of the most compelling applications of the \scc{hyphy} network is in forward modelling, where the underlying density field is reconstructed from observed data through an optimization process. For this application, the \scc{hyphy} network could replace analytical or semi-analytical approximations, such as the fluctuating Gunn Peterson approximation for Lyman alpha forest (e.g. in TARDIS,  \cite{2019TARDIS,2021TARDISII}). Our model is fully differentiable, allowing propagation through a model using first or second order methods. This is similar in spirit to work done in \citet{2018Chirag}, where a neural network was used to paint halo fields onto forward modeling dark matter density. However, this assumed a deterministic mapping from dark matter to galaxy light, while in \scc{hyphy} this mapping is controlled by a latent space. In this approach, hydrodynamical uncertainties could be marginalized out via sampling of the latent space during optimization, or jointly optimized for and then marginalized out via variational methods. 

In our loss function, we are implicitly assuming our latent space posterior can be well approximated by the dimensionality of our multivariate Gaussian latent space. It is possible a more accurate posterior would be achievable with an adversarial loss function which doesn't rely on this assumption, but such an approach comes with additional cost of difficulty in training. In addition, there is the added possibility of biasing the cosmological results due to the adversarial function implicitly learning the cosmology of the training set. For example, the adversarial loss function could implicitly learn to calculate a power spectra-like function, which would force the generative network to always produce maps with the the same power spectra as the training samples. With an L1 norm we are hopeful that our network would be transferable to other cosmological models, as these models would effect only the dark matter field without appreciably altering the hydrodynamical mapping.


We expect the overall design of the network to be easily extended to other hydrodynamical properties. Simulations suites such as CAMELS \citep{2020arXiv201000619V}, which include parametric feedback models, would allow the creation of an all-purpose mapping tool from dark matter to hydrodynamical models conditioned on underlying physics and redshift. While in \scc{nyx} the main source of effective stochastic processes are gas shocks, other hydrodynamical simulation tools like AREPO \citep{springel2010pur,2020AREPO}, used in CAMELS, allow for feedback processes through star formation, active galactic nuclei, supernova, etc. When studying these phenomena, it becomes very important to include a stochastic component to the network as done in \scc{hyphy}. We plan on further exploring these properties in future works.

\section*{Acknowledgments}
We thank Mustafa Mustafa, Stephanie Ger, Chirag Modi, Francios Lanusse, Peter Melchior, Uros Seljak, Charles Williams, Stanley Cox, and Earl Stevens for their helpful insights. We dedicate this work to the late Andre Louis Hicks, whose work is a constant inspiration. BH is supported by the AI Accelerator program of the Schmidt Futures Foundation.

This work was partially supported by the DOE's Office of Advanced Scientific Computing Research and Office of High Energy Physics through the Scientific Discovery through Advanced Computing (SciDAC) program.
This research used resources of the National Energy Research Scientific Computing Center, a DOE Office of Science User Facility supported by the Office of Science of the U.S. Department of Energy under Contract No. DEC02-05CH11231.

\appendix

\section{Lightcone Generation}
\label{app:lightcone}

\begin{figure*}
    \centering
    \includegraphics[trim={1cm 1cm 1cm 1cm},clip,width=0.950\textwidth]{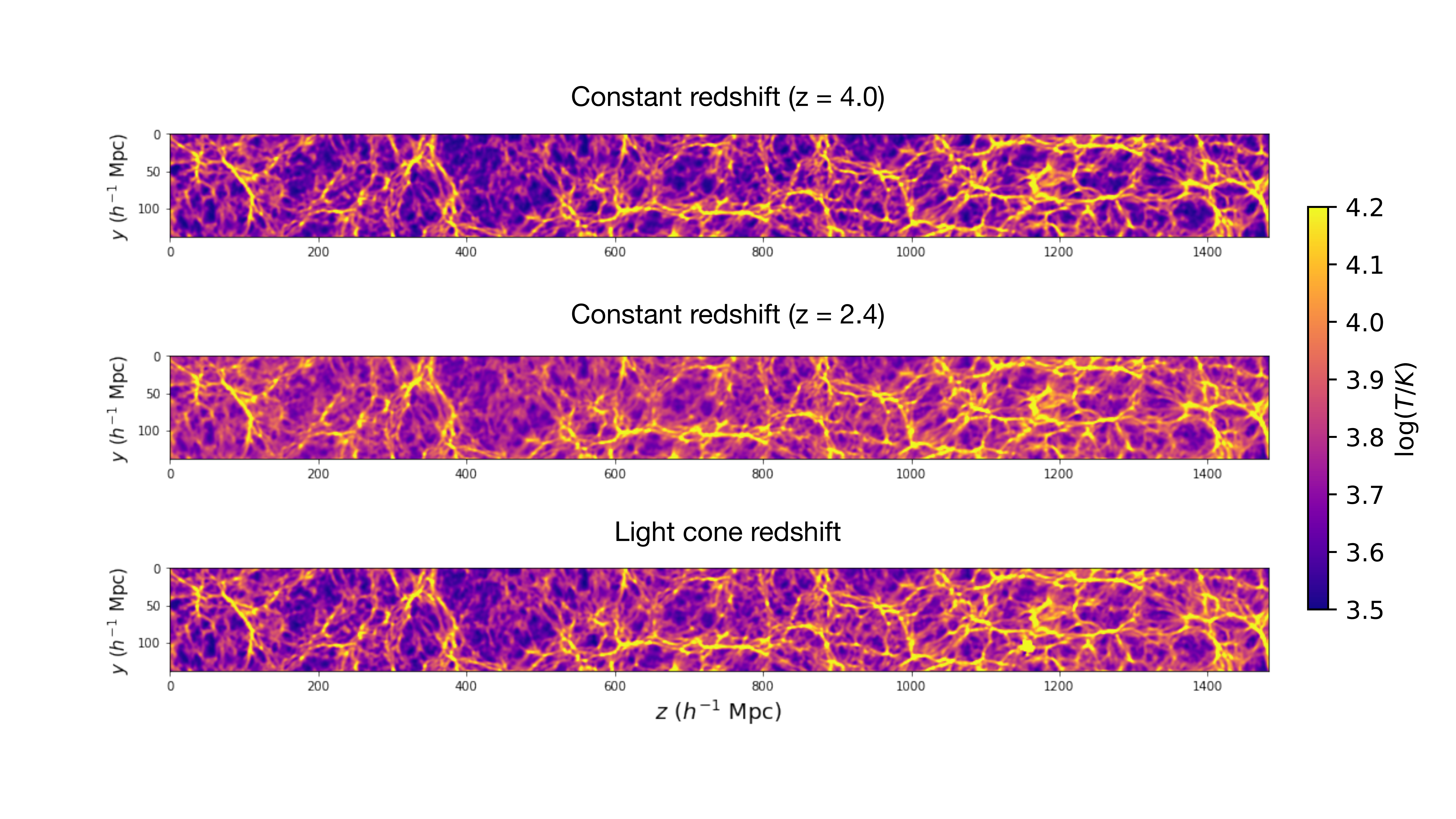}
    \caption{Figure demonstrating the generation of a light cone using \scc{hyphy}. In the top two panels, we show the \scc{hyphy} generated temperature corresponding to the same dark matter distribution as though it at $z=2.4$ and $z=4.0$. In the bottom panel we show the same distribution but with a varying redshift across the box, going from $z=4.0$ (left) to $z=2.4$ (right).}
    \label{fig:lc}
\end{figure*}
We construct the inputs to the \scc{hyphy} model to have a redshift label for every pixel, a necessary property to allow one model to learn a range of redshifts and maintain a fully convolutional architecture. While this is sub-optimal in terms of memory usage in the examples in the main body of the paper, it allows generation of realistic light-cones by varying this index over the box. As the convolutional layers are limited in scope, when mapping a given pixel they will only have input from a very limited range of redshifts. We therefore expect to not generate any significant artifacts due to the redshift variation over the test boxes that isn't present in the training sample. Our training sample only has fixed redshift boxes at $z=2.4$, $z=3.0$ and $z=4.0$. 

In practice the redshift label is most relevant for the temperature of the IGM; for a given fixed dark matter distribution there is little redshift dependence in the baryon density or velocity field. We demonstrate the construction of such a light cone in Figure \ref{fig:lc}. In the top two panels there is a clear difference in the overall temperature of the IGM with the temperature rising from $z=4.0$ to $z=2.4$. The lightcone version in the bottom panel, where we feed a varying redshift index, smoothly interpolates between the two. We plan to further apply this lightcone technique for mock generation and forward model reconstructions in future works.

\bibliographystyle{apj}

\bibliography{sample}

\end{document}